\documentclass[amsmath,amssymb,article,onecolumn]{revtex4}

\usepackage[english]{babel}
\usepackage{natbib}
\usepackage{multirow}

\usepackage{rotating} 
\usepackage{color}

\begin{document}

\title{Optimized structure and vibrational properties by error affected potential energy surfaces\footnote{
Reprinted  (adapted) with permission from J. Chem. Theory Comput., 2012, 8 (11), pp 4204--4215, DOI:10.1021/ct300576n. Copyright 2012 American Chemical Society.}}

\author{Andrea Zen}
\email{andrea.zen@uniroma1.it}
\affiliation{Dipartimento di Fisica, La Sapienza - Universit\`a di Roma, P.le Aldo Moro 2, 00185 Roma, Italy}

\author{Delyan Zhelyazov}
\affiliation{Dipartimento di Matematica Pura ed Applicata, Universit\`a degli Studi dell'Aquila, Via Vetoio 2, 67100 L'Aquila, Italy}
\affiliation{Centrum Wiskunde \& Informatica, Science Park 123, 1098 XG Amsterdam, the Netherlands}

\author{Leonardo Guidoni}
\email{leonardo.guidoni@univaq.it}
\affiliation{Dipartimento di Scienze Fisiche e Chimiche, Universit\`a degli Studi dell'Aquila, Via Vetoio 2, 67100 L'Aquila, Italy}

\begin{abstract}
The precise theoretical determination of the geometrical parameters of molecules at the minima of their potential energy surface and of the corresponding vibrational properties  are of fundamental importance for the interpretation of vibrational spectroscopy experiments. Quantum Monte Carlo techniques are correlated electronic structure methods promising for large molecules, which are intrinsically affected by stochastic errors on both energy and force calculations, making the mentioned calculations more challenging with respect to other more traditional quantum chemistry tools. To circumvent this drawback in the present work we formulate the general problem of evaluating the molecular equilibrium structures, the harmonic frequencies and the anharmonic coefficients of an error affected potential energy surface. The proposed approach, based on a multidimensional fitting procedure, is illustrated together with a critical evaluation of systematic and statistical errors. We observe that the use of forces instead of energies in the fitting procedure reduces the the statistical uncertainty of the vibrational parameters by one order of magnitude.
Preliminary results based on Variational  Monte Carlo calculations on the water molecule demonstrate the possibility to evaluate geometrical parameters, harmonic and anharmonic coefficients at this level of theory with an affordable computational cost and a small stochastic uncertainty ( $<0.07\%$ for geometries and $<0.7\%$ for vibrational properties).
\end{abstract}

\maketitle

\section{Introduction}


The accurate theoretical calculation of geometrical parameters at minima and of the corresponding vibrational properties are of great importance in Chemistry and Materials Science since they can help the interpretation of vibrational spectroscopy experiments such as infrared and Raman.
The general computational approach is to calculate relaxed geometries and vibrational frequencies from energies and derivatives of the potential energy surface (PES) obtained, within the Born-Oppenheimer approximation, via different quantum mechanical techniques such as Hartree-Fock, Density Functional Theory (DFT), Complete Active Space Self-Consistent Field (CASSCF), M{\o}eller-Plesset perturbation theory (PT), and Coupled-Cluster (CC).
Two different issues may currently limit the capability of quantum chemistry tools to quantitative predict vibrational properties of large molecules where electron correlation plays an important role: the accuracy of the computational method used and the complications introduced by the anharmonicity of the PES.
In the harmonic approximation, vibrational frequencies are calculated at the PES local minima by diagonalization of the mass-weighted Hessian matrix through normal mode analysis. An empirical way to consider the effects of anharmonicity consists in the introduction of "ad hoc" scaling factors \cite{HESS:1986wj,Scott:1996wc} to match the experimental frequencies.
More accurate results are obtained going beyond the harmonic approximation. Fully variational methods exists\cite{Carter:1997wr,Koput:2001uo,CassamChenai:2003gy}, but they are computationally affordable only for small molecules, 
while already  
for medium size molecules some approximations are necessary, still giving very accurate  results\cite{
Yagi:2004jy,Chaban:1999ur,Wright:2000tu,Wright:2000vt,Gregurick:2002fn,GERBER:1988ti,Bowman:1986wj,
Matyus:2007eb,Furtenbacher:2006bg,Czako:2004iq,
Krasnoshchekov:2008cp,Barone:2005p24347,Barone:2004id,Ruden:2003gx,Neugebauer:2003kb,Christiansen:2003do,Miani:2000uh,Dressler:1997vb,ALLEN:1990vw,SCHNEIDER:1989wl,CLABO:1988wn,Bailleux:1997wa}. 
Among these approximative methods, it is successfully widely used, an approach based on second order vibrational perturbation theory (PT2)\cite{Krasnoshchekov:2008cp,Barone:2005p24347,Barone:2004id,Ruden:2003gx,Neugebauer:2003kb,Christiansen:2003do,Miani:2000uh,Dressler:1997vb,ALLEN:1990vw,SCHNEIDER:1989wl,CLABO:1988wn,Bailleux:1997wa}, which takes the force fields of the PES around the configurational minimum up to the fourth order expansion to predict most of the spectroscopic parameters. An alternative approach, which directly includes  finite temperature nuclear motions, is represented by ab-initio molecular dynamics simulations\cite{B924048A,Bovi:2011gu},
where IR and Raman spectra can be directly obtained from the Fourier transform of dipole and polarizability autocorrelation functions.

Provided that we can proficiently account for the anharmonicity of the PES within a PT2 approach, at least for not too large-amplitude motions, the accuracy of the adopted electronic structure calculations becomes the leading ingredient for obtaining accurate results. A convenient balance between accuracy and computational cost is given by DFT, that scale as ${\cal O}(N_e^{2-3})$, being $N_e$ the number of electrons in the system.
However, well known drawbacks of the available exchange-correlation functionals \cite{Cohen:2008fg} limit the application of DFT to molecules where electronic correlation does not play a crucial role, leaving the correct treatment of such cases to more accurate post-Hartree-Fock methods. Unfortunately, the latter methods are computationally highly demanding  as the number of  electrons $N_e$ increases, since they scale as ${\cal O}(N_e^{5-7})$.

An alternative to the most traditional quantum chemical approaches is represented by  Quantum Monte Carlo (QMC)\cite{Austin:2012kt,Needs:2010p28123,Assaraf:2007gn,Foulkes:2001p19717}, a name that stands for a collection of  wave-function-based methods for the stochastic solution of the electronic many-body Schr\"odinger equation.
The simplest QMC approach is the Variational Monte Carlo (VMC), which relies on the evaluation of the expectation value of the molecular hamiltonian on a given parametrized wave function $\phi_T$. The parameters are chosen to minimize the variational energy which is calculated through stochastic integration. The choice of the wave function is crucial and is generally composed by an antisymmetrical determinantal part, aimed to describe static correlation effects, and a bosonic part, the Jastrow factor \cite{Jastrow:1955en}, which recovers most of the dynamical correlation effects.
Recent wave function developments include the Jastrow- antisymmetrized geminal power (J-AGP)\cite{Casula:2003p12694}, the Pfaffian\cite{Bajdich:2008p18507,Bajdich:2006p18510}, the backflow\cite{Holzmann:2003p28608}, and others multi-determinant-Jastrow functions\cite{Clark:2011ie,Fracchia:2012el}. 
Other QMC schemes beyond the variational ansatz are often based on projection methods. Although they are usually more accurate   they are computationally more demanding. For instance, the diffusion Monte Carlo (DMC) \cite{Foulkes:2001p19717} consists in an imaginary time  evolution of a trial wave function that in this way is projected onto the ground state.

Within QMC the energy of correlated molecular systems can be calculated with an accuracy in line with experimental data and with the most accurate traditional quantum chemistry methods \cite{Barborini:2012iy,Filippi:2012hg,Kolorenc:2011hv,Maezono:2010ic,Valsson:2010p25419,Spanu:2009p12613,Zimmerman:2009hh,Sterpone:2008p12640,Sorella:2007p12646,Schautz:2004p25285,Caffarel:1993td} but with a computational cost scaling with the size of the system as ${\cal O}(N_e^{3-4})$ \cite{Coccia2012JCC}.
This favorable scaling properties, the high accuracy of the results, and the fact that QMC algorithms are intrinsically parallel (demonstrated by an almost perfect scalability up to tens of thousands of processors on recent High Performance Computing facilities \cite{Coccia2012JCC}) should make QMC an optimal choice for accurate calculations of molecular geometries and vibrational properties.
On the other side, the underlying stochastic nature of the Monte Carlo algorithms is reflected into the fact that QMC energies and forces are affected by a stochastic error, which slowly decreases as the square root of the computational time. Due to this major drawback structural optimizations of molecules using QMC methods have been efficiently implemented only in the very last years\cite{Chiesa:2005p28489,Wagner:2010p25393,Sorella:2010p23644,Barborini:2012iy,Coccia2012JCC}. 
To tackle this problem in a recent work Sorella and coworkers \cite{Sorella:2010p23644} proposed a combination of the reweighting method\cite{Attaccalite:2008p12639}, the correlated sampling technique \cite{Filippi:2000p25406} and the space warp coordinate transformation\cite{UMRIGAR:1989tq,Assaraf:2003gq,Casula:2004p12689}. With the help of the algorithmic adjoint differentiation all the components of the ionic VMC forces can be calculated with and without pseudopotentials in a computational time that is only about four times that of an ordinary energy calculation\cite{Sorella:2010p23644}.
Vibrational properties, which require the calculation of second derivatives of the energy, to the best of our knowledge have never been calculated at full QMC level for systems larger than diatomic molecules\cite{Chiesa:2005p28489}. The accurate evaluation of relaxed molecular geometries using error affected potential energy surfaces is therefore still a challenge for computational chemistry and the establishment of an efficient procedure can be a breakthrough for the calculation based on QMC methods.

In this work we have developed a framework to estimate optimized geometrical parameters, harmonic frequencies and anharmonic constants of molecules where the potential energy surface is affected by a statistical uncertainty such as in the case of QMC calculations.
The goal has been achieved using a multi-dimensional fitting procedure of the PES and its first derivatives with respect to the ionic positions in the neighborhood of the minimum. A crucial role in the fitting procedure is represented by the choice of the number and positions of  grid points around the expected configurational minimum. The effects of the grid and of the mesh size have been investigated with the help of  preliminary DFT calculations where an uncertainty of the order of magnitude  of typical  QMC measures has been introduced.
In the first section of the present paper we review the calculation of vibrational properties from PT2, and in the second section we address the problem of fitting the PES using error affected measures. Finally we illustrate how to proceed in the choice of the grid points for the specific case of the water molecule demonstrating the feasibility of the method for QMC calculations.


\section{Calculation of Vibrational Properties}

We assume, as usual, to separate the electronic and nuclear part of the Coulomb Hamiltonian following the Born-Oppenheimer approximation.
We considere a set of configurations of the nuclear coordinates  close to the equilibrium structure, and for each configuration we obtain the ground state energy and nuclear forces using a DFT or QMC calculation.
The energy values can be fitted with a parametric function, in order to obtain the potential energy surface (PES) of the molecule in the neighborhood of the equilibrium structure.
This parametric PES enters into the nuclear hamiltonian, which describes the quantum mechanical motion of the nuclei, {\em i.e.} translations, rotations and vibrations of the molecule.

In the following section we will provide a short overview of the well established theory for the description of molecular vibrational properties, considering firstly the formulation of the nuclear problem in cartesian and internal coordinates, followed by the harmonic approximation, and secondly the second order perturbation theory (PT2) for the calculation  of the anharmonic corrections.
In the next section we will consider the problem of fitting the PES from QMC calculations, that differently from most quantum chemical methods for electronic structure calculation is affected by  stochastic errors.

\subsection{ Nuclear Hamiltonian }

The nuclear hamiltonian $\hat H$ is the summation of the kinetic $\hat T$ and potential $\hat V$ energy operators, the latter operator being given by the fitted PES. 
Given a molecule with $N$ atoms, the two terms can be easily written in terms of the $3N$-dimensional cartesian coordinates $\bf R$:
\begin{equation*}
\hat H = \hat T + \hat V 
  = -\frac{1}{2} \sum_\xi^N \sum_\alpha^3 \frac{1}{M_\xi} \frac{\partial^2 }{\partial R_{\xi,\alpha}^2}
    + V({\bf R}) \,,
\end{equation*}
where $M_\xi$ is the mass of the atom $\xi$ and $R_{\xi,\alpha}$ is its cartesian coordinate along the  $\alpha$-axis. The use of atomic units is always implicitly assumed hereafter.

The potential energy is generally assumed to be invariant under rotations and translations of the molecule as a whole (hypothesis of homogeneous and isotropic space), therefore  it is independent on the position of the center of mass and on the orientation of the molecule ({\em i.e.} the 6 {\em external} degrees of freedom for a non-linear molecule, or 5 in case of a linear one), and the potential energy is often conveniently written in terms of $3N-6$ (or $3N-5$ for a linear molecule) {\em internal} coordinates. 
For simplicity of notation, hereafter we will consider always the case of a non-linear molecule, so we will have  3N-6 internal  degrees of freedom and 6 external ones.  
%
The internal coordinates describe the vibrational motion of the molecule, and the external coordinates the rotational and translational motion.
The separation between an 
intramolecular term and a roto-translational term of motion, that is straightforward for the potential,  is  more complex for the kinetic part. 
In 1935 Carl Eckart showed, considering a body-fixed reference frame which rotates and translates with the molecule, that the factorization of the nuclear wave function in a vibrational and rotranslational one is never exact, as there are always non-vanishing vibration-rotation coupling terms\cite{Eckart:1935gk}. 
However there are some conditions, known as Eckart conditions, or Sayvetz conditions\cite{Sayvetz:1939hm}, which minimize these coupling terms.
A nuclear kinetic operator that includes these coupling term was obtained by Wilson and Howard\cite{Wilson:1936jj,Wilson:1936hf,Wilson:1937dz} in 1936, and refined by Darling and Dennison\cite{Darling:1940es} in 1940, until in 1968  Watson\cite{Watson:1968cm}  simplified it to obtain the form generally used nowadays, which is referred as Watson kinetic operator.
In the following sections we first review the harmonic approximation, where the coupling terms are completely neglected, and later a second order perturbation theory calculation, where coupling terms appears, coming from the simplified Watson form of the kinetic energy operator.

A molecule can usually be assumed semi-rigid, therefore its potential energy is proficiently written in terms of displacements from the equilibrium structure, corresponding to the (local) minimum of the PES. 
Given the equilibrium structure in cartesian coordinates $\bf R_0$; the cartesian displacement is ${\bf X} = {\bf R} -{\bf R}_0 $.

Often it is more convenient to use internal coordinates $\bar {\bf S}$ ({\em e.g.} valence coordinates: bond lengths, angles and dihedral angles) in place of the cartesian ones.
Indicating with $\bar {\bf S}_0$  the internal coordinates for the equilibrium structure that correspond to ${\bf R}_0$, a generic displacement from equilibrium in internal coordinates is 
${\bf S} = \bar{\bf S} -\bar{\bf S}_0$.
A mapping between the cartesian displacements $\bf X$ and the internal coordinates $\bf S$  exists, but in general this relation is linear only in first approximation, since generally internal coordinates  
are curvilinear. Close to the equilibrium structure we can Taylor expand $\bf S$:
\begin{equation}
S_i = B_i^a X_a + \frac{1}{2!} B_i^{ab} X_a X_b + \frac{1}{3!} B_i^{abc} X_a X_b X_c + \ldots
\label{eq_SX}
\end{equation}
where  $i=1,...,(3N-6)$ indexes the internal coordinates;  
$a,b,c=1,...,3N$  index the cartesian coordinates ({\em i.e.} components $x,y,z$ for each atom).
The Einstein summation notation of repeated indices is assumed hereafter, unless otherwise indicated or if in an expression summation terms are explicitly written.  
The coefficients in the series are the derivatives with respect to the Cartesian displacements:  
$B_i^{a} \equiv \frac{\partial S_i}{\partial X_a}$, 
$B_i^{ab} \equiv \frac{\partial^2 S_i}{\partial X_a \partial X_b}$, 
and so on.
The coefficients $B_i^{a}$ appearing in the linear term coincide with the well known Wilson $\bf B$ matrix \cite{Wilson:1955uh}.

The potential energy function for the nuclei in the proximity of the equilibrium structure is well described as a Taylor expansion in terms of the internal displacements coordinates:
\begin{equation}
V(S) = F + F^i S_i + \frac{1}{2!} F^{ij} S_i S_j 
       + \frac{1}{3!} F^{ijk} S_i S_j S_k
       + \frac{1}{4!} F^{ijkl} S_i S_j S_k S_l + \ldots
\label{eq_V(S)}
\end{equation}
where the coefficients in the expansion are defined as $F \equiv V$,
$F^i \equiv \frac{\partial V}{\partial S_i}$,
$F^{ij} \equiv \frac{\partial^2 V}{\partial S_i \partial S_j}$, \ldots
all calculated in $S=0$.
Clearly, if the configuration corresponding to $S=0$, {\em i.e.} the reference structure, is a minimum for the potential energy, the coefficients $F^i$ are all zero for the Euler conditions.
The coefficient $F$ represents the (local) minimum value of the potential energy, and it is an irrelevant offset for any vibrational calculation. 
For this reason it is usually ignored, considering implicitly $V(S)-F$ in place of $V(S)$.

\subsection{ Harmonic Approximation and Normal Modes}

The vibrational normal modes calculation in harmonic approximation is a standard approach, well known in literature\cite{Wilson:1955uh} and implemented in most  quantum chemistry codes.  
Here we will summarize the main results, that we will use below.

If it is  assumed that the reference structure is a minimum for the PES,
the linear term in the Taylor expansion of $V$ is zero. 
The main approximation in the harmonic approach is that only the leading terms, {\em i.e.} the quadratic (harmonic) ones, are considered in both the potential and kinetic energy, indicated hereafter by $V_\text{har}$ and $T_\text{har}$, and all remaining terms are neglected.
In internal coordinates the approximated potential and kinetic energy are respectively given by: 
\begin{equation}
V_\text{har}(S) = \frac{1}{2} F^{ij} S_i S_j \label{eq_Vh(S)}
\end{equation}
\begin{equation}
T_\text{har}(\dot S) = \frac{1}{2} (G^{-1})^{ij} \dot S_i \dot S_j \label{eq_Th(S)}
\end{equation}
where $\dot S_i$ is the time derivative of $S_i$, 
and  the 
$(3N-6) \times (3N-6)$ matrix $\bf G$ is calculated, according to Wilson\cite{Wilson:1955uh}, as:
\begin{equation*}
G^{ij} = \sum_{\xi}^{N} \sum_\alpha^{3} \frac{1}{M_\xi} B^{\xi,\alpha}_i B^{\xi,\alpha}_j 
\end{equation*}
being the coefficients $B_i^{\xi,\alpha}$ the same that appear in the linear term of (\ref{eq_SX}), upon  remapping of the indexes from $(\xi,\alpha)$ to $a$.

Introducing the $(3N-6)$ 
normal coordinates $Q_r$ it is possible to express both the potential and kinetic energies in diagonal form: 
\begin{equation}\label{eq_Vh(Q)}
V_\text{har}({\bf Q}) = \frac{1}{2} \sum_r \lambda_{r} (Q_r)^2 
\end{equation}
\begin{equation}\label{eq_Th(Q)}
T_\text{har}(\dot {\bf Q}) = \frac{1}{2} \sum_r (\dot Q_r)^2 
\end{equation}
where the coefficients $\lambda_r = 4\pi^2 c^2 {\omega_r}^2$ are the harmonic force constants, that provide  the harmonic frequencies of vibration $\omega_r$.
Once written in terms of the normal coordinates the problem is straightforward both in the classical and in the quantum case, since it  factorizes in  $(3N-6)$ independent harmonic oscillators (at least in case of nondegenerate frequencies, that we will not consider in this work). 

Assuming that normal coordinates $\bf Q$ are linearly related with internal coordinates $\bf S$ by the relation 
$S_i =  L_i^r Q_r$,  substituting  in (\ref{eq_Vh(S)}) and (\ref{eq_Th(S)}) and comparing with (\ref{eq_Vh(Q)}) and (\ref{eq_Th(Q)}) yields the relations for the transformation $L_i^r$, that written in matrix form are:
\begin{equation} \label{eqs_Lh}
{\bf L^+ F L} = \Phi
\quad \text{and} \quad
\bf L^+ G^{-1} L = 1
\end{equation}
where $\Phi$ is a diagonal matrix with elements  the harmonic force constants $\lambda_r$, $\bf 1$ is the $(3N-6)$ dimensional unit matrix, and  symbol $^+$ indicates the transpose matrix.
The second equation implies that 
$\bf L^+ = L^{-1} G$,
which  substituted in the first  condition yields: 
${\bf GFL = L} \Phi$.
Clearly, if we are only interested in the harmonic frequencies, it is enough to solve the secular equation: 
$\det({\bf GF}-\lambda{\bf 1}) = 0$.

\subsection{ Anharmonic Corrections by Second Order Perturbation Theory }

Using second order perturbation theory (PT2)\cite{Hoy:1972ee,MillsRao:1972,Herman:1999vc,Barone:2005p24347,Krasnoshchekov:2008cp}  
it is possible to go beyond the  harmonic approximation and to take into account  anharmonic effects,  expanding  the PES to the third and forth order terms.

The first point in the approach is to consider the forth order 
potential energy expansion written in terms of normal coordinates $\bf Q$:
\begin{equation} \label{eq_V(Q)}
V({\bf Q}) = \frac{1}{2!} \Phi^{rs} Q_r Q_s
     + \frac{1}{3!} \Phi^{rst} Q_r Q_s Q_t
     + \frac{1}{4!} \Phi^{rstu} Q_r Q_s Q_t Q_u 
\end{equation}
where we recognize in $\Phi^{rs} = \lambda_r \delta_{rs}$ the harmonic force constants already introduced in the previous section. 
The coefficients 
$\Phi^{rst} \equiv \frac{\partial^3 V}{\partial Q_r \partial Q_s \partial Q_t}$ and 
$\Phi^{rstq} \equiv \frac{\partial^4 V}{\partial Q_r \partial Q_s \partial Q_t \partial Q_q}$ are the anharmonic force constants, respectively of third and forth order.

Assuming that we have fitted the PES in terms of the internal displacement coordinates $\bf S$ around the minimum structure, obtaining the values of the tensors $F$ that appear in (\ref{eq_V(S)}) (the zeroth and first order terms are clearly irrelevant), we need to map this values into the tensors $\Phi$ that appear in (\ref{eq_V(Q)}).
To do this we need to consider the transformation from internal curvilinear coordinates $\bf S$ to normal rectilinear $\bf Q$. Differently from what we did in the harmonic approach, here we must consider that this relation is not linear, since the internal coordinates are curvilinear.
We can consider the following Taylor expansion of $\bf S$ in terms of $\bf Q$:
\begin{equation} \label{eq_SQ}
S_i = L_i^r Q_r + \frac{1}{2!} L_i^{rs} Q_r Q_s + \frac{1}{3!} L_i^{rst} Q_r Q_s Q_t + \ldots
\end{equation}
where the transformation elements are called $L$ tensor elements\cite{Hoy:1972ee}.
The linear terms $L_i^r$ coincide with the matrix $\bf L$ in (\ref{eqs_Lh}) which solves the secular equation in the harmonic calculation.
Substitution of (\ref{eq_SQ}) in (\ref{eq_V(S)}) and comparison with (\ref{eq_V(Q)}) provides the relation between the $\Phi$ coefficients and the $F$ ones:
\begin{eqnarray}
\Phi^{rs} &=& {F}^{ij}L^{r}_{i}L^{s}_{j} \\
{\Phi}^{rst} 
 &=& {F}^{ijk}L^{r}_{i}L^{s}_{j}L^{t}_{k} +
     {F}^{ij} ( L^{rs}_{i} L^{t}_{j} + 
                L^{rt}_{i} L^{s}_{j} + 
                L^{st}_{i} L^{r}_{j} ) \\
{\Phi}^{rstu} 
 &=& {F}^{ijkl} L^{r}_{i} L^{s}_{j} L^{t}_{k} L^{u}_{l} + \nonumber \\
&& + {F}^{ijk} ( L^{rs}_{i} L^{t}_{j} L^{u}_{k} + 
                 L^{rt}_{i} L^{s}_{j} L^{u}_{k} + 
                 L^{ru}_{i} L^{s}_{j} L^{t}_{k} +
                 L^{st}_{i} L^{r}_{j} L^{u}_{k} + 
                 L^{su}_{i} L^{r}_{j} L^{t}_{k} + 
                 L^{tu}_{i} L^{r}_{j} L^{s}_{k} ) + \nonumber \\
&& + {F}^{ij} ( L^{rs}_{i} L^{tu}_{j} + 
                L^{rt}_{i} L^{su}_{j} + 
                L^{ru}_{i}L^{st}_{j}) 
   + {F}^{ij} ( L^{rst}_{i} L^{u}_{j} +
                L^{rsu}_{i} L^{t}_{j} +
                L^{rtu}_{i} L^{s}_{j} + 
                L^{stu}_{i} L^{r}_{j} ) 
\end{eqnarray}
in agreement with \cite{Hoy:1972ee}.
%
The $L$ tensor elements can be determined using the following relations: 
\begin{eqnarray}
L^{r}_{i} &\equiv& \frac{\partial S_{i}}{\partial Q_{r}} 
   = \frac{\partial S_{i}}{\partial X_{a}} 
     \frac{\partial X_{a}}{\partial Q_{r}} 
   = B^{a}_{i} H^{r}_{a}\\
L^{rs}_{i} &\equiv& \frac{\partial^2 S_{i}}{\partial Q_{r}\partial Q_{s}}
   = \frac{\partial^2 S_{i}}{\partial X_{a}\partial X_{b}}
     \frac{\partial X_{a}}{\partial Q_{r}}
     \frac{\partial X_{b}}{\partial Q_{s}} 
   = B^{a b}_{i}H^{r}_{a}H^{s}_{b}\\
L^{r s t}_{i} &\equiv& 
     \frac{\partial^3 S_{i}}{\partial Q_{r}\partial Q_{s}\partial Q_{t}}
   = \frac{\partial^3 S_{i}}{\partial X_{a}\partial X_{b}\partial X_{c}}
     \frac{\partial X_{a}}{\partial Q_{r}}
     \frac{\partial X_{b}}{\partial Q_{s}}
     \frac{\partial X_{c}}{\partial Q_{t}}
   = B^{a b c}_{i}H^{r}_{a}H^{s}_{b}H^{t}_{c}
\end{eqnarray}
where the coefficients
$ H^r_a \equiv \frac{\partial X_a}{\partial Q_r} $
can be calculated, using  a matrix notation,  as 
$ {\bf H} = {\bf M^{-1} B^+ (L^{-1})^+ } $, and 
the tensors $B^{ab}_{i}$ and $B^{abc}_{i}$ can be obtained from $\bf B$ by numerical differentiation. 


Once all the coefficients of the expansion (\ref{eq_V(Q)}) are determined, we consider the Watson Hamilton operator up to the fourth order of expansion in normal coordinates $Q$, that  is\cite{MillsRao:1972,Herman:1999vc,Barone:2005p24347} (the Einstein convention is not used from here to the rest of the section):
\begin{equation}
\hat{H} = 
   \frac{1}{2} \sum_r \left( \hat{P_r}^2 + \lambda_{r} \hat{Q_{r}}^2 \right) 
   + \frac{1}{6}  \sum_{rst}  \Phi^{rst}  \hat{Q_r} \hat{Q_s} \hat{Q_t}
   + \frac{1}{24} \sum_{rstu} \Phi^{rstu} \hat{Q_r} \hat{Q_s} \hat{Q_t} \hat{Q_u}
   + \sum_{\alpha} B^{e}_{\alpha} {j_{\alpha}}^2
\end{equation}
in which:
\begin{equation}
j_{\alpha} = 
   \sum_{r < s} \zeta_{rs}^{\alpha} 
   \left(  \hat{Q_r} \hat{P_s}
        -  \hat{Q_s} \hat{P_r} \right)
\end{equation}
In these expressions 
$\hat Q_r$ is the operator relative to $r^{th}$ normal coordinate, and 
$\hat P_r \equiv -i \frac{\partial}{\partial Q_r}$ is the corresponding momentum operator, conjugate to $\hat Q_r$;
$\alpha=1,2,3$ represents the three rotational axes in Eckart frame,  
$B^{e}_\alpha = \frac{1}{2 I_{\alpha}}$ the corresponding equilibrium rotational constants, 
being $I_{\alpha}$ the principal moments of inertia;
$\zeta_{rs}^{\alpha}$ represents the Coriolis coupling constants.

The second order perturbative calculation provides the coefficients in the  expansion for the molecular vibrational energy:
\begin{equation}
G({\bf n}) = \chi_0 +
  \sum_r \omega_r \left( n_r +\frac{1}{2} \right) + 
  \sum_{r \le s} \chi_{rs} \left( n_r +\frac{1}{2} \right) \left( n_s +\frac{1}{2} \right) 
\label{eq_Gn}
\end{equation}
where ${\bf n} = \{ n_r \}$ are the $(3N-6)$ quantum numbers associated to corresponding vibrational modes.
In particular the anharmonic coefficients $\chi_{rs}$ are calculated as\cite{MillsRao:1972}:
\begin{equation} \label{eq_Xrr}
\chi_{rr} = 
   \frac{1}{16} \frac{ \Phi_{rrrr}}{{\omega_r}^2}
 - \frac{1}{16} 
      \sum_{s} \frac{ {\Phi_{rrs}}^2 }{ {\omega_r}^2 {\omega_s}^2}
       \frac{ 8{\omega_r}^2 -3{\omega_s}^2 }{(4{\omega_r}^2-{\omega_s}^2)}
\end{equation}
and for $r\ne s$:
\begin{eqnarray} \label{eq_Xrs}
\chi_{rs} &=& 
   \frac{1}{4} \frac{\Phi_{rrss}}{\omega_r \omega_s} - 
   \nonumber \\
&&- \left\{ 
    \frac{1}{4} \sum_{t} 
      \frac{ {\Phi}_{rrt} {\Phi}_{sst} }{\omega_r \omega_s {\omega_t}^2} 
   -\frac{1}{2} \sum_{t} 
      \frac{{\Phi_{rst}}^2}{\omega_r \omega_s }
      \frac{ {\omega_r}^2 + {\omega_s}^2 - {\omega_t}^2 }
             {\left( (\omega_r+\omega_s)^2-{\omega_t}^2 \right)
              \left( (\omega_r-\omega_s)^2-{\omega_t}^2 \right)}
   \right\} + 
   \nonumber \\
&&+ \sum_{\alpha} 
     B_{\alpha}^{e} (\zeta_{rs}^{\alpha})^2
     \left(
       \frac{\omega_r}{\omega_s} + \frac{\omega_s}{\omega_r}
     \right)
\end{eqnarray}
where the first term in  $\chi_{rr}$ and in $\chi_{rs}$ 
comes from the first order perturbation, and all others come from the second order perturbation. The last term in  $\chi_{rs}$  
accounts for the contribution of the Coriolis coupling. 
These formulae are valid in the case of no degenerate harmonic frequencies and no quantum-mechanical resonances; for details on these specific cases see for instance references \citenum{Krasnoshchekov:2008cp,Barone:2005p24347,Hoy:1972ee}.

Equation (\ref{eq_Gn}) yields the expressions for 
the fundamental frequency $\nu_r$:
\begin{equation} \label{eq_fundamental_freq}
\nu_r = \omega_r + \Delta_r 
      = \omega_r + 2 \chi_{rr} + \frac{1}{2} \sum_{s\ne r} \chi_{rs}
\end{equation}
overtones: 
\begin{equation}
[2\nu_r] = 2 \omega_r + 6 \chi_{rr} + \sum_{s\ne r} \chi_{rs}
         = 2 \nu_r + 2 \chi_{rr} 
\end{equation}
combination bands:
\begin{equation}
[\nu_r \nu_s] = \omega_r + \omega_s + 2 \chi_{rr} + 2 \chi_{ss} + 2 \chi_{rs} 
                + \frac{1}{2} \sum_{k\ne r,s} (\chi_{rk}+\chi_{sk})
         = \nu_r + \nu_s + \chi_{rs} 
\end{equation}
%


\section{Fit of the PES by error affected measures}

\subsection{ Difficulties due to error affected measures }
 
The methodology introduced in the previous sections for the calculation of vibrational  frequencies requires tight (for the harmonic approach) or very tight (for PT2) geometry optimization criteria\cite{Johnson:2010fy,Barone:2004id,Barone:2005p24347}, 
namely the configuration at  minimum should have a residual gradient smaller than $10^{-5}$ or $10^{-7}$~a.u, respectively. 
The consequence of an inaccurate knowledge of the structural minimum would indeed propagate on the frequencies, yielding to unreliable results, especially for PT2 calculations.
These tight optimization criteria are easily achieved using most common electronic structure methods like  the self consistent field (SCF) cycles used in DFT and post-Hartree-Fock methods.

Unfortunately in QMC calculations the situation is drastically different, because every QMC measure is affected by a stochastic error $\sigma$ that is inversely proportional to the square root of the number of sampling $N_s$, that in turn is proportional to the computational time required for the calculation. This means that to reduce the error of a factor 10 it would be necessary a computational effort 100 times larger. At opposite, in a SCF calculation this would just cost few additional SCF iterations. For this reasons QMC stochastic errors on energies are typically much larger then what desirable for vibrational frequencies calculations, being of the order or $\sigma_E \sim 10^{-4}$~a.u. for the energy evaluation and ,$\sigma_F \sim 10^{-3}$~a.u. for each force component. It is important to point out that, thanks to recent developments in QMC obtained by Sorella and Capriotti \cite{Sorella:2010p23644}, the calculation of all force components is 
only four times more computationally demanding then a single energy calculation.

Before introducing the formalism of the proposed fitting procedure we  discuss  some general issues about the localization of a minimum using an error affected PES.  Let's suppose we would like to locate the minimum of the PES having information on energies affected by a stochastic error.
From a structural point of view, if we want to find the distance $\Delta$ from the minimum in a potential that is approximatively $E={1 \over 2}k \Delta^2$ using a single point estimation of the energy $E$ affected by an error $\sigma_E$, it yields that $ |\Delta| = \sqrt{2 E / k}$ is affected by an error $\sigma_\Delta \sim {\sigma_E /( k \Delta )}$. As example, considering the OH bond stretching force constant $k\sim 0.5$a.u., with an error on the energy of $\sigma_E \sim 10^{-4}$a.u., it is clear that it is not possible to estimate $\Delta$ with an accuracy larger than $\sim 1.4*10^{-2}$a.u. 
The situation is even worst for angles, since the bond bending force constants are typically one order of magnitude lower than the stretching constants.
The computational effort to resolve the position difference $\Delta$ is therefore proportional to $1/ (k^2 \Delta ^4 )$, which clearly becomes prohibitive for a tight (small $\Delta$) structural optimization.

A different situation comes out if the information on forces, and not only on energies, is known. Close to the minimum the force is approximatively $F=-k \Delta$, therefore  $\Delta$ can be calculated from a single point evaluation of the force as $\Delta=-F/k$, and the associated error is $\sigma_\Delta = \sigma_F / k$. 
Considering again the previous example of the OH bond stretching, with a typical $\sigma_F \sim 10^{-3}$a.u. we have an accuracy on $\Delta$ of the order $2*10^{-3}$a.u., not enough for a tight optimization but much better than the previous energy-based estimation. In this case the computational effort to resolve a difference $\Delta$ using forces is therefore proportional to $1/(k^2 \Delta^2)$, much more convenient than the case of energy.

\subsection{Multidimensional Fit of PES}

Taking into account the previous considerations, in our approach the equilibrium position and the other parameters of the PES near a minimum are calculated using a multidimensional fit of independent data points (energies or forces) calculated on a grid  around an initial guess of the structural minimum of the molecule.

The chosen parametric function ${\cal V}_0$ that describes the PES in the proximity of the minimum $\bar {\bf S}_0$ is the Taylor expansion of the potential, see (\ref{eq_V(S)}), cut to the fourth order term.
Therefore the parameters to fit are  
${\bf K}=  \{  {\bar{S}_0}^i, F,F^{ij},F^{ijk},F^{ijkl}\}$, 
where  the first order coefficients $F^i$ are not included as they have to be zero in the minimum.
As a result of the fitting procedure we will have an estimate of the minimum position  ${\bar{S}_0}^i$, the hessian matrix $F^{ij}$, 
and the 3rd and 4th order coefficients for the PT2 calculation of the anharmonic corrections.

The fitting function ${\cal V}_0$ presents a nonlinear dependence on the parameters of the structural minimum ${\bar{S}_0}^i$.
As will appear clear later on, for technical reasons, it would be more convenient to use a linearized form of this function instead of all its parameters. 
In the following we therefore introduce a self-consistent approach, which is easier to implement but is completely equivalent.
Let's consider a different fitting function ${\cal V}$  that represents the 4th order expansion of the PES around a structure 
${\bar{S}_G}^i$ and is only a guess of the exact minimum.
As a consequence of this, the linear coefficients $F^i$ in the expansion can not be assumed equal to zero any more, in fact they will be as small as  ${\bar{S}_G}^i$ is close to ${\bar{S}_0}^i$.
Therefore we obtain the equation:   
\begin{equation}
{\cal V}(\textbf{S}_G,\textbf{k}_G) = F_G 
       + F_G^i S_G^i 
       + \frac{1}{2} F_G^{ij} S_G^i S_G^j 
       + \frac{1}{6} F_G^{ijk} S_G^i S_G^j S_G^k
       + \frac{1}{24} F_G^{ijkl} S_G^i S_G^j S_G^k S_G^l 
\label{eq_VG(S)}
\end{equation}
that is linearly parametrized by
${\bf k}_G=\{ F_G,F^i_G,F^{ij}_G,F^{ijk}_G,F^{ijkl}_G \}$
and functionally depends on the displacement
${\bf S}_G = \bar{\bf S} - {\bar{\bf S}_G}$
from the guess configuration.

The fitted coefficients ${\bf k}_G$ are relative to the guess structure ${\bar{\bf S}_G}$, and can not be used directly for the harmonic and PT2 calculation. However, once we have fitted the function, it is easy to obtain a new guess ${\bar{\bf S}_{G(2)}}$ of the minimum configuration, just finding the configuration that has the gradient of the fitted potential equal to zero, and then to repeat the fitting process of  ${\cal V}({\bf S}_{G(2)},{\bf k}_{G(2)})$ with respect to this new guess structure ${\bar{\bf S}_{G(2)}}$, obtaining a new set of coefficients ${\bf k}_{G(2)}$.
Since the fitting process is computationally very fast, we can easily iterate this procedure a sufficiently large number of times, converging at the end to the equilibrium configuration $\bar {\bf S}_{G(n)} \to \bar {\bf S}_0$ and obtaining the fitting parameters ${\bf k}_{G(n)} \to {\bf k}_0$ that represents the coefficients around the real minimum. 

At each iteration $n$ we have to identify the configuration that minimize the gradient, in order to obtain the new guess structure $\bar {\bf S}_{G(n+1)}$.
The exact solution is not so straightforward, since the gradient of the potential has terms of 2nd and 3rd order in the displacement, however within the described iterative scheme the solution to the leading linear order is enough to ensure the over mentioned convergence.
Indicating with ${\bf H}_{G(n)}$ the matrix formed by the elements of $F_{G(n)}^{ij}$, and with ${\bf F}_{G(n)}$ the vector given by $F_{G(n)}^i$, the approximated condition of zero gradient is written as: 
$ {\bf F}_{G(n)} + {\bf H}_{G(n)} \cdot ( \bar{\bf S}_{G(n+1)} - {\bar{\bf S}_{G(n)}} ) = {\bf 0} $
that yields:
\begin{equation}\label{eq_iter_S_G}
\bar{\bf S}_{G(n+1)}  = \bar{\bf S}_{G(n)} 
                      - {{\bf H}_{G(n)}}^{-1} \cdot {\bf F}_{G(n)} 
\end{equation}
being ${{\bf H}_{G(n)}}^{-1}$ the Moore-Penrose pseudoinverse of ${{\bf H}_{G(n)}}$.
In our calculations we have seen that a few ($n \sim 3$) iterations are enough to obtain a tight convergence ({\em i.e.}  residual forces $F_{G(n)}^{i} \ll 10^{-10}$a.u.).
In agreement with (\ref{eq_iter_S_G}), zero residual forces means that we can confidently take $\bar{\bf S}_{G(n)}$ as the minimum structure.

In this work we report several results, namely the coefficients of the PES, the harmonic frequencies and the anharmonic corrections, that depend on the error affected measurements of energies and forces.
The error propagation on the results have  been obtained using a standard resampling method, namely the Jackknife variance estimator\cite{KUNSCH:1989ws,Wolff:2004cu}.
%
In the following two subsections we will see how the values of the fitting parameters are inferred using respectively error affected energy or force measures.
For simplicity of notation the subscript $_G$ will not be reported, as it is clear that the explanation concerns a single step in the iteration, with displacements $\bf S$ and fitting values $\bf k$ relative to a particular guess structure. 

\subsection{Fit of PES using  Energy  measures } \label{section_fit_energy}

We want to show here how to determine the values of the parameters $\bf k$ of the function 
${\cal V}({\bf S},{\bf k})$
that fits the error affected energy measures around the minimum.
Consider a set of $N_\alpha$ configurations 
$\{ \textbf{S}^\alpha \}$, for 
$\alpha = 1,\ldots, N_\alpha$.
For each configuration $\textbf{S}^\alpha$ the calculations provide a measure of the potential energy 
$\tilde V(\textbf{S}^\alpha) \equiv \tilde V_\alpha$ 
and a corresponding stochastic error\cite{FLYVBJERG:1989wl} 
$\sigma_{\tilde V}(\textbf{S}^\alpha) \equiv \sigma_\alpha$. 

The likelihood that a particular $\textbf{k}$ fits the QMC energy measures 
$D_E = \{ \textbf{S}^\alpha, \tilde V_\alpha , \sigma_\alpha \}_{\alpha=1,\ldots,N_\alpha}$ is given by the function\cite{Jaynes:2003ua}:
\begin{equation}\label{eq_L(k|E)}
{\cal L}({\bf k} | D_E) = p(D_E|\textbf{k}) =
  \prod_\alpha^{N_\alpha} { 1 \over \sqrt{2\pi {\sigma_\alpha}^2 } } 
    \exp\left\{ - { \left[ {\cal V}(\textbf{S}^\alpha,{\bf k}) - \tilde V_\alpha  \right]^2
                  \over 2 {\sigma_\alpha}^2 } \right\}
\end{equation}
where we have assumed that independent estimations of the energy at each configuration are normally distributed. Actually for VMC energy estimations, using the VMC standard sampling technique, this is true only in the limit of infinite sampling, while for finite sampling heavy-tails coming from the local energy singularities are observable as reported by J.R. Trail\cite{Trail:PRE2008_1}. 
The estimation of these leptokurtotic tails has unknown bias and should be evaluated for each individual case, and their magnitude can be reduced increasing the VMC sampling. We have explicitly verified that in our particular system the sampling we have used is enough to make negligible the deviations from a normal distribution. 
For the sake of completeness it should be addressed that neither when reweighting sampling techniques are used the probability density distribution for the estimated energy is normal\cite{Trail:PRE2008_2}.

It is not necessary for our purposes to obtain the complete distribution of the likelihood function in terms of $\bf k$, but it is enough to find the particular value of $\textbf{k}_M$ that maximizes the likelihood.
It is convenient to consider the logarithm of (\ref{eq_L(k|E)}), using the fact that the logarithm is a monotone transformation, and after dropping out some irrelevant terms it yields that $\textbf{k}_M$ is the minimum of the chi-squared function:
\begin{equation}
{\chi_E} ^2 (\textbf{k}) = 
  \sum_\alpha^{N_\alpha} 
    \left( \frac{ {\cal V}(\textbf{S}^\alpha,{\bf k}) - \tilde V_\alpha }{\sigma_\alpha} \right)^2
\end{equation}

A remarkable property of the used fitting parameters 
${\bf k}=\{ F,F^i,F^{ij},F^{ijk},F^{ijkl} \}$
is that the fitting function depends linearly on them, therefore:
\begin{equation}\label{eq_linV(Sk)}
{\cal V}(\textbf{S},{\bf k}) 
  \equiv 
    \sum_p^{N_p} k^p {\cal V}_p(\textbf{S})
\end{equation}
where $p$ is an index running over all the $N_p$ parameters $\bf k$, and 
${\cal V}_p(\textbf{S}) \equiv \frac{ \partial {\cal V}(\textbf{S},{\bf k})}{\partial k^p}$
is not functionally dependent on $\bf k$.

By substituting  (\ref{eq_linV(Sk)}) in (\ref{eq_L(k|E)}) it can be seen that the stationary condition 
$\frac{\partial {\chi_e}^2  }{\partial k^p} = 0$
is equivalent to solve:
$\textbf{A} \cdot \textbf{k}_M - \textbf{B} = \textbf{0}$, 
where the elements of the $N_p$ dimensional square matrix $\bf A$ are:
\begin{equation}
A_{pq} = \sum_\alpha (\sigma_\alpha)^{-2}
  {\cal V}_p(\textbf{S}^\alpha) {\cal V}_q(\textbf{S}^\alpha)
\end{equation}
and the $N_p$ dimensional vector $\bf B$ is:
\begin{equation}
B_{p} = \sum_\alpha (\sigma_\alpha)^{-2}
  {\cal V}_p(\textbf{S}^\alpha) 
  \tilde V_\alpha
\end{equation}
It follows that the fitting parameters are  
$ {\bf k}_M = \textbf{A}^{-1} \textbf{B} $,
where $\bf A^{-1}$ is the Moore-Penrose pseudo-inverse of matrix $\bf A$.

\subsection{Fit  of PES using  Force measures}  \label{section_fit_force}

The methodology used for fitting the PES using force measures is very similar to the one used for the energy, but there is a little complication due to the fact that the forces come from the derivative of the potential with respect to the Cartesian coordinates $\textbf{R}$, not the curvilinear coordinates $\textbf{S}$ used for the fitting function ${\cal V}({\bf S},{\bf k})$.

There is a many-to-one mapping between the cartesian coordinates $\textbf{R}$ and the corresponding curvilinear coordinates $\textbf{S} = \textbf{S}(\textbf{R})$, because whereas given $\textbf{R}$ the internal coordinates $\textbf{S}(\textbf{R})$ are univocally defined, the opposite is not true due to the arbitrariness in the choice for $\textbf{R}(\textbf{S})$ of the global orientation of the molecule.
Since our QMC code works in cartesian coordinates, we have decided to use $\textbf{R}$ also for this fitting.
It follows that the cartesian force fitting function of component $a$, with $a=1,\ldots,3N$, is:
\begin{equation}\label{eq_Fa(Rk)}
{\cal F}^a(\textbf{R},{\bf \tilde k}) 
   = - \frac{\partial {\cal V} (\textbf{S}(\textbf{R}),{\bf \tilde k})}{\partial R_a}
   = - \sum_i^{3N-6} B_i^a(\textbf{R}) 
     \frac{\partial {\cal V} (\textbf{S}(\textbf{R}),{\bf \tilde k})}{\partial S_i}  
\end{equation}
where 
$ B_i^a(\textbf{R}) \equiv \frac{\partial S_i(\textbf{R})}{\partial R_a} 
$  
is given by the Wilson $\bf B$ matrix, calculated for the reference position $\textbf{R}$.
%
The fitting parameters  
$\tilde{\bf k}=\{ F^i,F^{ij},F^{ijk},F^{ijkl} \}$
differ from the set $\bf k$ in the energy fitting only for the absence of the $F$, that has been dropped out in the derivative, see equation~(\ref{eq_Fa(Rk)}).
The number of parameters $\bf \tilde k$ in the fit with the force are therefore $N_p -1$, indexed with $p=2,\ldots,N_p$.
%
It is clear from  (\ref{eq_linV(Sk)}) and (\ref{eq_Fa(Rk)})  that ${\cal F}^a(\textbf{R},{\bf \tilde k})$ is a linear function of $\bf \tilde k$, therefore we can write:
\begin{equation}\label{eq_linF(Sk)}
{\cal F}^a(\textbf{R},{\bf \tilde k}) 
  =  \sum_{p=2}^{N_p} k^p {\cal F}^a_p(\textbf{R})
\end{equation}
where the terms
$
{\cal F}^a_p(\textbf{R}) 
  \equiv 
    \frac{\partial {\cal F}^a (\textbf{S}(\textbf{R}),\tilde{\bf k})}{\partial k^p}
$
are all independent from the parameters $\bf k$,
and by substituting in (\ref{eq_Fa(Rk)}), these terms can be easily calculated as: 
$
{\cal F}^a_p(\textbf{R}) 
  = - B_i^a(\textbf{R}) 
     \frac{\partial {\cal V}_p (\textbf{S}(\textbf{R}))}{\partial S_i}  
$.

For the fitting we consider a set of $N_\alpha$ configurations 
$\{ \textbf{R}^\alpha \}$, for 
$\alpha = 1,\ldots, N_\alpha$, and 
for each configuration we have an associated force measure
$\tilde F^a(\textbf{R}^\alpha) \equiv \tilde F^a_\alpha$ 
with an associated stochastic error
$\sigma_{\tilde F^a}(\textbf{R}^\alpha) \equiv  \sigma^a_\alpha$, 
being $a=1,\ldots,3N$ the component index.
When the force components come from a QMC calculations, they are usually correlated since they are derived by the same stochastic random walk. In this case also their $3N$-dimensional variance-covariance matrix $\textbf{C}_\alpha$, which has on the $a$th diagonal elements the square of $\sigma^a_\alpha$, can be calculated.

The likelihood function for the parameters $\tilde{\bf k}$ to fit the QMC measures 
$D_F = \{ \textbf{R}^\alpha , \tilde {\bf F}_\alpha , \textbf{C}_\alpha \}_{\alpha=1,\ldots,N_\alpha}$
is given by the function:
\begin{equation}
{\cal L}(\tilde{\bf k} | D_F) = p(D_F|\tilde{\bf k}) =
  \prod_\alpha^{N_\alpha} { 1 \over (2\pi)^{3N/2} \sqrt{\det(\textbf{C}_\alpha)} }   
    \exp\left\{ - {1\over 2}  \sum_{a,b}^{3N} 
      { ({\textbf{C}_\alpha}^{-1})_{ab}  
        \left[ {\cal F}^a(\textbf{R}^\alpha,\tilde{\bf k}) - \tilde F^a_\alpha  \right]  
        \left[ {\cal F}^b(\textbf{R}^\alpha,\tilde{\bf k}) - \tilde F^b_\alpha  \right] 
      } 
    \right\}
\end{equation}
as follows from definition of likelihood, having assumed a normal distribution for the QMC force estimates. 
In some cases this assumption could be not exact, depending on the method adopted for the calculation of the forces and the finite VMC sampling\cite{Trail:PRE2008_1,Trail:PRE2008_2}.  In particular for the calculations reported in this paper we have verified that the deviations of the force estimations from a normal distribution are negligible

In the present work we did not estimated the full variance-covariance matrix, but only the diagonal part, therefore afterwords  we will neglect the correlation between the force components, and we will use only the standard deviations $\sigma^a_\alpha$ of each force component.

As in the case of the energy fitting, we only need to maximize the likelihood function, that is equivalent to finding the $\tilde{\bf k}_M$ that minimize the ${\chi_F} ^2$ function:
\begin{equation}
{\chi_F} ^2 (\tilde{\bf k}) = 
  \sum_\alpha^{N_\alpha} \sum_a^{3N}
    \left( \frac{ {\cal F}^a(\textbf{R}^\alpha,{\bf \tilde k}) - \tilde F^a_\alpha }{\sigma^a_\alpha} \right)^2
\end{equation}
Again the stationary configuration is given by the condition 
$ \frac{\partial {\chi_F} ^2}{\partial k^p} = 0 $,
that is equivalent to solve $ \tilde{\bf A} \cdot \tilde{\bf k}_M = \tilde{\bf B}$, 
where the elements of the $N_p -1$ dimensional square matrix $\bf \tilde A$ are:
\begin{equation}
\tilde A_{pq} = 
  \sum_\alpha^{N_\alpha} \sum_{a}^{3N} 
    (\sigma_\alpha^a)^{-2}
    {\cal F}_p^a(\textbf{R}^\alpha) 
    {\cal F}_q^a(\textbf{R}^\alpha)
\end{equation}
and the $N_p -1$ dimensional vector $\bf \tilde B$ is:
\begin{equation}
\tilde B_{p} = 
  \sum_\alpha^{N_\alpha} \sum_{a}^{3N} 
    (\sigma_\alpha^a)^{-2}
    {\cal F}_p^a(\textbf{R}^\alpha) 
    \tilde F_\alpha^a
\end{equation}
for $p,q = 2,\ldots, N_p$.
It follows that 
$ \tilde {\bf k}_M = \tilde {\bf A}^{-1} \cdot \tilde {\bf B} $,
where $\bf \tilde A^{-1}$ is the Moore-Penrose pseudo-inverse of matrix $\bf \tilde A$.

\subsection{The choice of the fitting grid}\label{section_grid}

The choice of the grid points where energies and forces are evaluated is a crucial aspect for the proposed approach. 
Clearly the larger is the region of the PES explored by energy or force measurements, the lower is the influence of the stochastic errors on the parameters of the function fitting the PES, and ultimately on the equilibrium structure and the vibrational frequencies.
%
On the other side, the parametric fitting function is an expansion where we are neglecting all the terms beyond the fourth order expansion, therefore the larger the region of the PES expansion, the lower the accuracy of the fitting function and the larger the systematic error.
This implies that from one side we want to consider the larger mesh size in order to reduce the effect of the stochastic errors on the fitted parameters, but on the other side the mesh can not be too large, otherwise we would introduce a systematic error due to an inaccurate fitting function in the range where the QMC measures were obtained.
To study this effects for typical values of stochastic errors we studied the behavior of the systematic and stochastic errors using grid of different sizes and shapes. We considered as model system for these tests the DFT PES of the water molecule where stochastic errors were introduced artificially on forces and energies. In this way we can directly compare the parameters obtained with the fitting procedure with the exact parameters of the PES to directly evaluate the effect of different grids.
In the specific case of the water molecule, the grid points are chosen as follows:
\begin{eqnarray*}
 r_1 &= r_0 + i *\Delta r  \\
 r_2 &= r_0 + j *\Delta r  \\
 \phi &= \phi_0 + k *\Delta \phi 
\end{eqnarray*}
where the center of the grid, given by $r_0$ and $\phi_0$, is our initial guess of the equilibrium structure.
The integers $i,j,k$ are so that
$ |i|,|j|,|k| \leq N_1 $;  
$ j \leq i $ ;
and  if $i=j$ then 
$|i|+|j|\leq N_2 $, 
otherwise $|i|+|j|+|k| \leq N_3 $.
Therefore each mesh is specified by the parameters $N_1,N_2,N_3, \Delta r, \Delta \phi$.
The set of meshes considered in this paper are reported in Table~\ref{table:grid}.

\begin{table}
\begin{tabular}{ c  c  c  c  c  c  c  }  
\hline
type   & $N_1$ & $N_2$ & $N_3$ & $\Delta r$[a.u.] & $\Delta \phi$[deg.] & \# points \\
\hline
mesh-1 &  2    & 3     & 4     & 0.02 &  1$^{\circ}$ & 59 \\
mesh-2 &  2    & 3     & 4     & 0.03 &  3$^{\circ}$ & 59 \\ 
mesh-3 &  2    & 3     & 4     & 0.04 &  5$^{\circ}$ & 59 \\
mesh-4 &  2    & 3     & 4     & 0.08 & 10$^{\circ}$ & 59 \\
mesh-5 &  2    & 3     & 4     & 0.20 & 20$^{\circ}$ & 59 \\

mesh-6 &  3    & 3     & 4     & 0.03 &  3$^{\circ}$ & 77 \\
mesh-7 &  4    & 4     & 5     & 0.03 &  3$^{\circ}$ & 139 \\
mesh-8 &  5    & 5     & 6     & 0.03 &  3$^{\circ}$ & 208 \\
\hline
\end{tabular}
\caption{ Mesh types used in this paper, that are univocally defined by the values of $N_1,N_2,N_3, \Delta r, \Delta \phi$. For a description of the construction method see Section~\ref{section_grid}.}
\label{table:grid}
\end{table}


\section{Computational Details}

In this work we have performed both DFT and QMC calculations: the first ones aimed to assess the optimal choice of the grid points around the configurational minimum to take under control both the systematic and the stochastic errors; the second to test the validity of the methods when applied to QMC calculations.

DFT calculations were performed using the {\em ORCA} package \cite{Neese:2005cy}, with the {\em B3LYP} functional\cite{STEPHENS:1994vd,Becke:1993vx,LEE:1988ub,Vosko:1980ui} and {\em aug-cc-pVTZ} basis set\cite{KENDALL:1992vx,DUNNING:1989uk}. For each point of the grids we calculate energies and gradients. To simulate the presence of stochastic errors we added to the values of energies and forces random contributions normally distributed with the desired standard deviation $\sigma_E$ and $\sigma_F$, respectively for the energy and the force. 

The QMC energy and force calculations have been carried out using the {\em TurboRVB} package developed by S. Sorella\cite{TurboRVB} that includes a complete suite of variational and diffusion quantum Monte Carlo programs for wave function and geometry optimization of molecules and solids.
The wave function used is a Jastrow Antisymmetrized Geminals Power\cite{Casula:2003p12694} (J-AGP). 
Since in the present context the use of QMC is purely demonstrative of the feasibility of the PES fitting procedure, we used a minimal wave function, in order to minimize the computational costs.
The orbital basis set for the AGP part is composed of $[4s,4p,1d]$ Gaussian type  orbitals (GTO) contracted in $(2s,2p,1d)$ for the oxygen atom, and of $[4s,1p]$ GTO orbitals contracted in $(2s,1p)$ for the hydrogen atoms. The exponent values were taken from the {\em cc-pVDZ} basis set, considering only values lower that 20 a.u. 
The Jastrow factor used here consists of several terms that account for the 1-body, 2-body, 3-body and 4-body interaction between electrons and nuclei, as described in refs. \citenum{Casula:2004p12689,Casula:2003p12694}.
For the 3/4-body Jastrow part we used a GTO basis set with $(1s,1p)$ uncontracted orbitals for oxygen and $(1s)$ for hydrogen.
An energy-consistent pseudopotential\cite{Burkatzki:2007p25447} was used for the oxygen atom.
The wave function optimization was carried on, for each ionic configuration, on the AGP matrix parameters, on the contraction coefficients of the atomic AGP bases, and on all the Jastrow parameters, including exponents. 
All the QMC calculations here presented have been obtained inside a Variational Monte Carlo (VMC) scheme, in which the VMC forces\cite{Casula:2004p12689} can be computed very efficiently using the adjoint algorithmic differentiation recently developed and implemented by S. Sorella and L. Capriotti\cite{Sorella:2010p23644}. 

The experimental configuration of the molecule was taken as the initial guess of the PES structural minimum.
We calculated the QMC forces for the experimental configuration, using the setup described above, and we observed that a residual force of the order of $10^{-3}$a.u. is present. This implies that the experimental configuration is close enough to the PES minimum to be taken as the center of the grid for the PES fitting that we are proposing here, but it would not be close enough for a single point evaluation of the hessian matrix and harmonic frequencies.



\section{Results and Discussion}
 
In this work we considered as test case for our approach the water molecule,
and specifically the most common $H_2 O$ molecule with $^1H$ and $^{16}O$ isotopes. 
Water counts for three internal coordinates, or nine cartesian coordinates.  
The considered internal coordinates are the two distances $r_1$ and $r_2$ between the oxygen and the hydrogens, and the HOH angle $\phi$.
In the equilibrium configuration 
experiments\cite{Benedict:1956id} reported  
$r_\text{EXP} = 0.95721(3)${\AA} and 
$\phi_\text{EXP} = 104.522(5)^\circ$.
State of the art computational results, obtained with coupled-cluster calculations, give equilibrium geometry quite close to the experimental ones\cite{Feller:2009p23440}, while an overview of the results that can be obtained by various quantum chemical computations are reported in ref. \citenum{KIM:1995p23441}.
The number of independent parameters in a $4^{th}$ order expansion of the potential in internal coordinates, taking into account the symmetry of the system, are 22 for the energy fit and 21 for the force fit.
The number of measures (of energies or forces) has to be larger than the number of parameters in order to have a well defined fitting problem.

In the following we will first investigate the problem of the systematics versus stochastic errors using DFT calculations, and finally we will report results obtained for a VMC calculation performed using a small wave function.

In order to investigate the systematic errors induced by the approximated parametrization of the PES, we consider here the  grids from mesh-1 to mesh-5 defined in Section~\ref{section_grid} and  Table~\ref{table:grid}.
By fitting the forces provided by DFT we obtained the  results reported in Table~\ref{table:anharm1}. The comparison with the analytical DFT parameters in the first column shows how for small meshes the computed frequencies and anharmonic corrections are close to the analytical results, but as the mesh size increases they start to differ. 
This happens because the polynomial expansion is a good approximation of the potential in a not too big neighborhood of the equilibrium configuration. From this calculation we can therefore estimate up to which mesh size the systematic errors can be negligible with respect other sources of errors. 
In particular for mesh-4 the systematic errors are less then
$\sim 6\mbox{cm}^{-1}$, {\em i.e.} $\sim 0.2\%$, on the frequencies
and less then $\sim 13\mbox{cm}^{-1}$ on the anharmonic coefficients.

\begin{table}
\begin{tabular}{ c  c  c  c  c  c  c  }  
\hline
 & {\em ORCA} & mesh-1  & mesh-2 & mesh-3 & mesh-4 & mesh-5 \\  
 & & $0.02B/1^{\circ}$  & $0.03B/3^{\circ}$ & $0.04B/5^{\circ}$ & $0.08B/10^{\circ}$ & $0.20B/20^{\circ}$ \\ 
\hline 
$\omega_2$ & 1626.3 & 1626.7 & 1626.6 & 1626.8 & 1628.9 & 1635.7\\ 
$\omega_1$ & 3793.4 & 3793.7 & 3794.1 & 3794.5 & 3798.7	& 3853.5 \\
$\omega_3$ & 3896.1 & 3896.3 & 3896.7 & 3897.1 & 3900.3 & 3953.6 \\
\hline
$X_{22}$  &         & -19.53 & -19.66 & -19.68 & -19.70 & -19.64 \\
$X_{12}$  &         & -13.41 & -13.20 & -13.27 & -13.91 & -20.43 \\
$X_{23}$  &         & -16.70 & -16.20 & -16.36 & -16.75 & -21.08 \\
$X_{11}$  &         & -44.25 & -44.42 & -44.83 & -47.63 & -70.39 \\
$X_{13}$  &         &-173.42 &-173.96 &-175.46 &-186.21 &-274.72 \\
$X_{33}$  &         & -50.31 & -50.73 & -51.07 & -53.90 & -77.42 \\ 
\hline
\end{tabular}
\caption{\textbf{ Systematic errors induced by the mesh size }
Harmonic frequencies $\omega$ and anharmonic corrections $X$ (both in $cm^{-1}$)  computed for a DFT calculation, for the {\em B3LYP} functional and {\em aug-cc-pVTZ} basis set. 
In the first column the calculation has been performed using directly the  {\em ORCA} package\cite{Neese:2005cy} (anharmonic coefficients are not reported because not provided by the  {\em ORCA} package), while
in the remaining columns we used the fitting the PES with forces in a grid with some different mesh types characterized by the $\Delta r$ and $\Delta \phi$ reported below the mesh number (see details in the text and in Table~\ref{table:grid}). }
\label{table:anharm1}
\end{table}

\subsection{Stochastic errors}
\begin{figure}[tb!]
\begin{center}
\includegraphics[width=.4\textwidth]{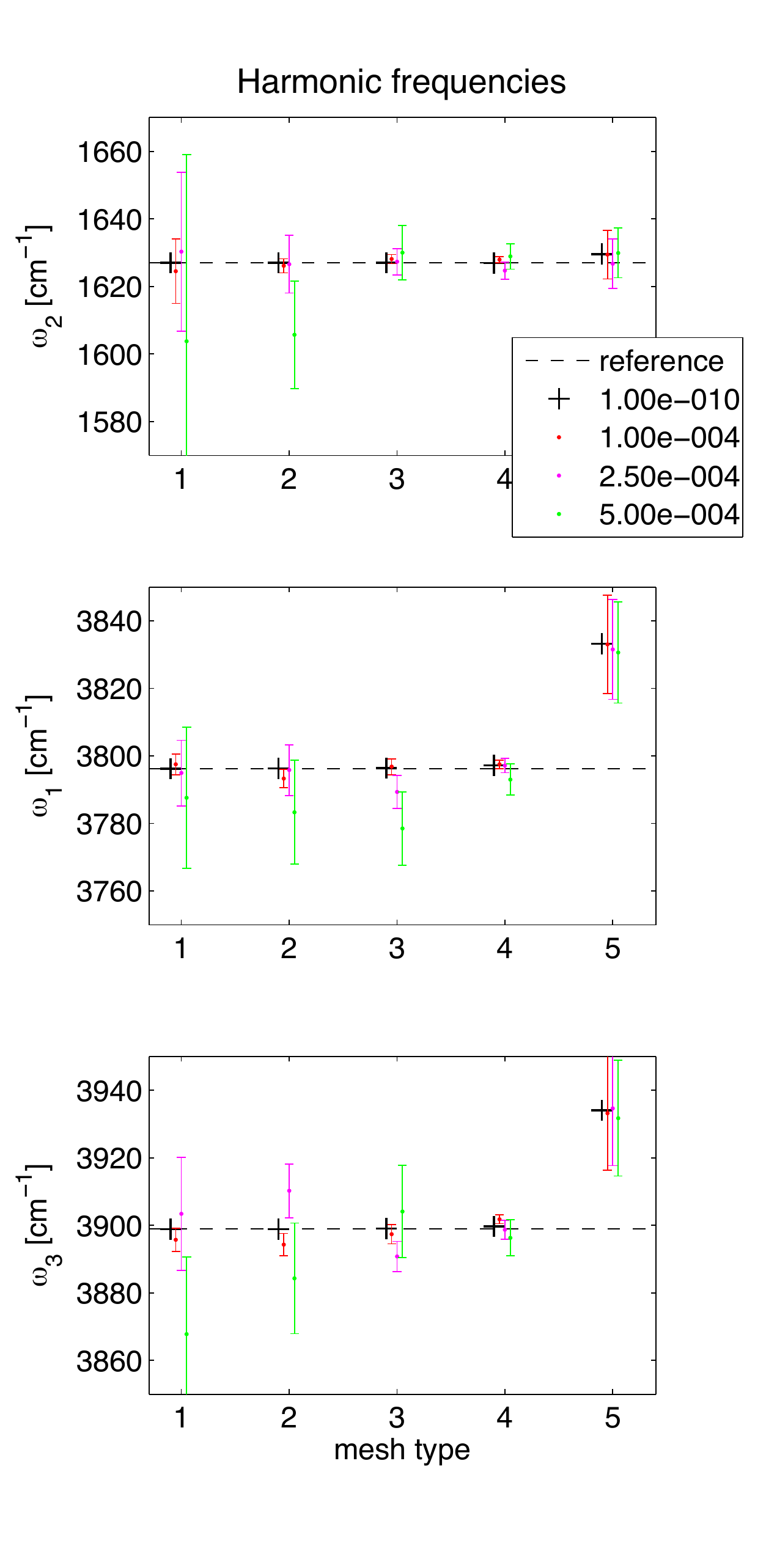}
\end{center}
\caption{ Harmonic frequencies obtained fitting DFT (B3LYP/aug-cc-pVTZ) forces, where random errors normally distributed have been added in order to simulate QMC stochastic errors  for different mesh types (details in the text). 
Several values (labels) of the standard deviation for the force components  where tried around the typical QMC ones.
For each case we reported e representative frequency with the corresponding error calculated using the Jackknife resampling method. 
The dashed line represents the reference results given by ORCA package, that can be considered free from any stochastic and grid error.
}\label{fig_omega}
\end{figure}

\begin{figure}[tb!]
\begin{center}
\includegraphics[width=.7\textwidth]{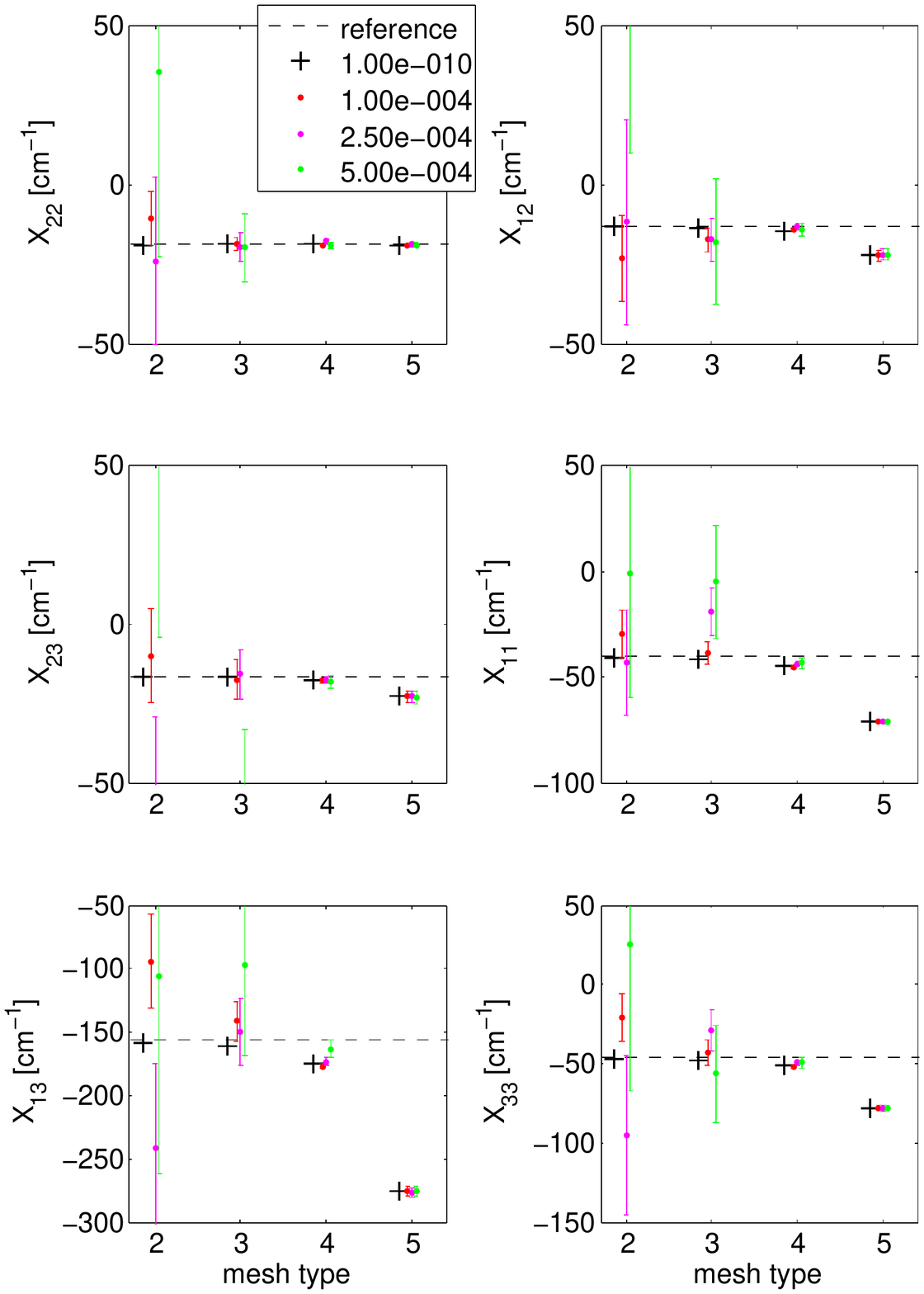}
\end{center}
\caption{ Same of the Figure~\ref{fig_omega}, but here the anharmonic force constants are reported. } 
\label{fig_X}
\end{figure}

To address the effects of the propagation of the stochastic error we consider again the grids from mesh-1 to mesh-5.
We use the DFT forces previously  computed, and to simulate the stochastic errors of the QMC forces  we add to each DFT force component a pseudorandom number normally distributed and with a variance $\sigma^2$. 
The results obtained fitting these forces are reported in Figures~\ref{fig_omega} and \ref{fig_X}, where the different colors correspond to different values of $\sigma$, and the dashed line is the reference result ({\em i.e.} tiny mesh and no stochastic errors). The vertical lines are the error bars. 
As the mesh size increases the stochastic errors decrease, however the systematic errors increase. We need to choose our mesh in order to have a good balance between stochastic and systematic errors for a typical value of the QMC $\sigma$.
Considering the stochastic errors of the forces having magnitude $\sigma = 5*10^{-4}$a.u., 
Figures~\ref{fig_omega} and \ref{fig_X} 
and Table~\ref{table:anharm1} suggest that mesh-4 has the best balance between stochastic and systematic errors.

Once established which is the optimal size of the grid, we can address here the importance of the way the points are distributed on the grid (the grid type) and of the influence of the total number of points on the grid.
For this purpose we consider the grids: mesh-6, mesh-7 and mesh-8, as defined in Table~\ref{table:grid}. The results of the fitting  procedure of the forces are reported in Tables~\ref{table:harmonic_results} and \ref{table:anharmonic_results}, 
respectively for the harmonic frequencies $\omega$ and the anharmonic corrections $X$.
We may observe, by comparison with  Table~\ref{table:anharm1}, that the systematic errors are under control in all cases, being 
$\leq 2 cm^{-1}$ for the harmonic frequencies and 
$\leq 5 cm^{-1}$ for the anharmonic corrections.

We would like now to compare different strategies for decreasing the stochastic errors in our computations, having a certain fixed amount of computer time.
Let $t$ be the computational time required to obtain certain stochastic errors of the harmonic frequencies $\sigma_{\omega}$, 
$N$ the number of points in the grid, 
and $n_s$ in the number of Monte Carlo samples in each single point calculation that provides forces with a stochastic error of $\sigma_f$. 
We have that $\sigma_\omega \propto \sigma_f / \sqrt{N}$, and 
$\sigma_f \propto 1 / \sqrt{n_s}$.
The total computational time $t$ is the sum of the time for the sampling $t_s\propto N*n_s$ and the time for the wave function optimization $t_{opt} \propto N$.
The more we sample the wave function, the more accurate has to be the optimization, therefore we can assume that the time for the optimization is a fraction $x_{opt}$ of the time for the sapling. It follows that
$t \propto (1+x_{opt})*n_s*N$.

The same amount of additional computer time can be distributed in different ways, either increasing the length of each stochastic sampling or increase the number of sampling points. The first strategy is to decrease the stochastic errors by longer Monte Carlo sampling, without adding additional points to the grid. The only quantity that varies is therefore $n_s$.
The second strategy is to add new points to the grid, namely changing $N$.

The numerical results in Table~\ref{table:grid} are carried out using grids spanning the same configurational region but using a finer ($N_{8}=208$ points), medium ($N_{7}=139$ points) or coarser ($N_{6}=77$ points) mesh.
According to our previous considerations, if we take fixed the $\sigma_f$ for each point calculation, the computational time for a QMC calculation is  $t_6 = c * N_6$ when using mesh-6  and $t_8 = c * N_8$ when using mesh-8, being $c$ an opportune constant.
Therefore the expected increment in the computational time is $t_8/t_6 \sim 2.7$.
The reduction of the stochastic error obtained in this way according to our simulations 
is $\sim 0.31(9)$ for the harmonic frequencies $\omega$, see Table~\ref{table:harmonic_results},
and $\sim 0.21(5)$ for the anharmonic constants $X$, see Table~\ref{table:anharmonic_results}.
These results have to be compared with the reduction that we would obtain  using the first strategy for the same increment of computational time, that is $\sqrt{t_6/t_8}\sim 0.6$ both for $\omega$ and $X$.
In conclusion the calculations reveal that to reduce the stochastic error on frequencies and anharmonic constant is more effective to add additional mesh points rather then increase their accuracy.

\begin{table}
\begin{tabular}{ c   c   c   c   c  }
\hline
& $\sigma_{F}$ & mesh-6 & mesh-7 & mesh-8\\
&              & 77 points & 139 points & 208 points \\

\hline
             & 1.00e-10 &1627(0) &1627(0) &1627(0)\\
$\omega_{2}$ & 5.00e-04 &1634(8) &1621(5) &1628(3)\\
             & 7.50e-04 &1612(15) &1618(7) &1633(6)\\
             & 1.00e-03 &1662(41) &1643(11) &1630(7)\\ 
\hline
             & 1.00e-10 &3795(0) &3795(0) &3796(0)\\
$\omega_{1}$ &5.00e-04 &3805(11) &3801(6) &3798(4)\\
             &7.50e-04 &3815(22) &3790(8) &3796(5)\\
             &1.00e-03 &3823(33) &3803(11) &3787(6)\\ 
\hline
              &1.00e-10 &3897(0) &3898(0) &3898(0)\\
$\omega_{3}$  &5.00e-04 &3920(12) &3890(6) &3899(4)\\
              &7.50e-04 &3894(16) &3911(9) &3891(6)\\
              &1.00e-03 &3902(20) &3915(13) &3907(8)\\ 
\hline
\end{tabular}
\caption{\textbf{ Harmonic frequencies from error affected measures. }
Harmonic frequencies $\omega$ (in $cm^{-1}$) for different mesh types (see description in the text) vs stochastic noise magnitude. 
Similarly to 
Figures~\ref{fig_omega} and \ref{fig_X} 
each reported frequency value is a single representative case, not the average frequency for the corresponding standard deviation. The error on each frequency is calculated using the Jackknife resampling.
Below the mesh number is reported the number of actual points in the grid.
}
\label{table:harmonic_results}
\end{table}

\begin{table}
\begin{tabular}{ c  c  c  c  c }
\hline
& $\sigma_{F}$ & mesh-6 & mesh-7 & mesh-8\\
&              & 77 points & 139 points & 208 points \\
\hline
          & 1.00e-10 &-20(0)  &-20(0)   &-20(0)\\
$X_{22}$  & 5.00e-04 &-39(11) &-18(5)   &-17(3)\\
          & 7.50e-04 &-10(20) &-11(9)   &-18(5)\\
          & 1.00e-03 &-24(24) &-27(10)  &-19(8)\\ 
\hline
           &1.00e-10 &-13(0)  &-13(0)   &-13(0)\\
$X_{12}$   &5.00e-04 &-65(40) &-23(19)  &-10(7)\\
           &7.50e-04 & 20(47) &-15(25)  &-31(10)\\
           &1.00e-03 &-40(86) &-14(37)  & -7(15)\\ 
\hline          
		   &1.00e-10 &-16(0)  &-16(0)   &-16(0)\\
$X_{23}$   &5.00e-04 &-67(49) &-31(18)  &-13(9)\\
		   &7.50e-04 &-62(68) &-22(32)  &-10(10)\\
		   &1.00e-03 & 51(83) &-67(31)  &-28(17)\\ 
\hline
		   & 1.00e-10 &-45(0)  &-45(0)  &-46(0)\\
$X_{11}$   & 5.00e-04 &-41(25) &-52(8)  &-44(4)\\
		   & 7.50e-04 &-71(26) &-50(8)  &-49(6)\\
		   & 1.00e-03 &-69(58) &-50(14) &-36(5)\\ 
\hline
		   &1.00e-10 &-176(0)  &-178(0)  &-178(0)\\
$X_{13}$   &5.00e-04 &-232(53) &-175(15) &-191(10)\\
		   &7.50e-04 &-124(52) &-177(29) &-156(13)\\
		   &1.00e-03 &-183(78) &-141(29) &-154(17)\\ 
\hline
		   &1.00e-10 & -51(0)  & -52(0)  &-52(0)\\
$X_{33}$   &5.00e-04 &-151(53) & -59(27) &-47(11)\\
		   &7.50e-04 &-100(75) & -88(31) &-41(16)\\
		   &1.00e-03 &  -3(90) &-158(38) &-82(18)\\ 
\hline
\end{tabular}
\caption{\textbf{ Anharmonic constants from error affected measures. }
Analogous to previous table, but this reports anharmonic constants $X$ (in $cm^{-1}$) for different mesh types vs stochastic noise magnitude.  }
\label{table:anharmonic_results}
\end{table}


To demonstrate the effectiveness of the method in a real albeit simple QMC case, we performed VMC calculations using a minimal basis sets for the AGP and Jastrow part, on a mesh-4 grid centered around the experimental structural minimum, {\em i.e.}
$r=0.9572${\AA} and $\phi=104.52^{\circ}$.
On this structure we obtained a VMC energy of $-17.2144(6)$a.u. and residual force components $\le 0.003(1)$a.u.
The forces calculated in the grid points spam from $-0.13$a.u. to $0.15$a.u., and have  an average stochastic error of $\sim 0.0009$a.u.

In  Table~\ref{tab:QMC_fit} are reported the results of two different fits: one using the VMC energies and the other using the VMC forces. It is immediately clear how the quantities obtained from the energy fit have stochastic errors which are at least one order of magnitude larger than those obtained by the fit of the forces. This trend is observed both for the evaluation of the geometrical parameters at minimum and for the harmonic and anharmonic constants. Since the calculations of QMC forces in our scheme requires only four times the computational effort of energy evaluations\cite{Sorella:2010p23644}, it is clear that the force-fitting procedure is always extremely more convenient with respect the energy-fitting procedure.

Despite the variational wave function used was very small, the  equilibrium configuration predicted using the  force fit yields to 
$r_\text{VMC} = 0.9552(2)${\AA}    and $\phi_\text{VMC} = 104.11(7)^{\circ}$, that are in good agreement with the experimental values, and perform well in comparison with the other computational methods\cite{KIM:1995p23441}. On the other hand, the harmonic frequencies are overestimated if compared to the experimental values, whereas the estimation of the anharmonic constants are rather close.
This discrepancy was not unexpected, because the basis sets for the AGP and Jastrow expansion are minimal. A complete basis set study requires much larger computational resources and will be the subject of further investigations.
We point out that besides the discrepancy with respect to the experimental results, the statistical errors on the estimated harmonic frequencies are as small as 0.7\%, demonstrating that the proposed method is able to provide for small molecules a precise evaluation of vibrational frequencies.

\begin{table}
\begin{tabular}{ c   c   c   c  }
\hline
  & energy-fit  & force-fit & EXP\\
\hline
$r$[\AA] 	& 0.954(3)   &	0.9552(2)  &  0.95721(3) \\
$\phi$[deg.]	& 104.4(6)   &	104.11(7)  &  104.522(5) \\ 
\hline
$\omega_2$ 	& 1637(99)   &	 1723(11)  &    1648.47  \\
$\omega_1$ 	& 3678(302)  &	 3920(14)  &    3832.17  \\
$\omega_3$ 	& 4197(322)  &	 4039(17)  &    3942.53  \\ 
\hline
$X_{22}$  	&  -11(47)   &	  -19(3)   &    -16.81   \\ 
$X_{12}$  	&  234(150)  &	  -21(6)   &    -15.93   \\ 
$X_{23}$  	&  -16(113)  &	  -22(7)   &    -20.33   \\ 
$X_{11}$  	&  -12(236)  &	  -46(6)   &    -42.57   \\ 
$X_{13}$  	& -975(537)  &	 -169(20)  &    -165.82  \\ 
$X_{33}$  	& -294(289)  &	  -49(10)  &    -47.57   \\ 
\hline
$\nu_{2}$ 	&  1724(66)  &	 1663(6)   &    1594.59  \\
$\nu_{1}$ 	&  3283(433) &	 3732(10)  &    3656.65  \\  
$\nu_{3}$ 	&  3113(465) &	 3846(13)  &    3755.79  \\ 
\hline
\end{tabular}
\caption{\textbf{ QMC results.}
Equilibrium configuration (OH distance $r$ in {\AA} and angle $\phi$ in degrees), harmonic frequencies $\omega$, anharmonic constants $X$ and fundamental frequencies $\nu$, 
theoretically computed by VMC energy fitting (first column), VMC force fitting (second column),  and experimentally observed by Benedict et al.
\cite{Benedict:1956id} 
(third column). 
For all the computational results the standard mesh $N_1=4, N_2=2, N_3=3$ was used, with $\Delta r = 0.08$~a.u. and $\Delta \phi = 10^{\circ}$ (see Section~\ref{section_grid}). 
The energy[force] was evaluated at each point of the mesh and the potential expansion up to 4$^{th}$ order was fitted as described in Section~\ref{section_fit_energy}[Section~\ref{section_fit_force}]. 
The numbers in the brackets are the stochastic errors, estimated by the jackknife algorithm.
}
\label{tab:QMC_fit}
\end{table}

 
\section{Conclusions}

In the present work we proposed a theoretical framework for the calculation of equilibrium structures and vibrational properties through a potential energy surface affected by stochastic noise. The proposed procedure has as natural application the QMC electronic structure calculations, although it is not limited to these methods.
We demonstrated that applying  a multidimensional fitting procedure to the QMC energies or forces around the minimum, we can obtain accurate results on the structural parameters at the minimum and on the vibrational frequencies, despite the presence of the intrinsic stochastic error of the method.
The theoretical developments in this work , as well as the numerical results, are aimed to minimize the computational time and mitigating the impact of the stochastic errors on the accuracy of the computed harmonic frequencies.
As QMC example, we applied our technique to the calculation of the equilibrium structure, harmonic frequencies and anharmonic constants of the water molecule. The obtained results demonstrated how using information of forces instead of energies it is possible to reduce the statistical error on the vibrational frequencies of at least one order of magnitude, leading to an average error lower than 0.07\% for the geometrical parameters and 0.7\% for the harmonic frequencies. 
These results open the possibility to apply Quantum Monte Carlo methods for the evaluation of vibrational spectroscopic parameters of molecules, although the scaling of the method with the number of atoms in the system limits the approach to relatively small molecules. 



\section*{Acknowledgement}
We thank Sandro Sorella and Matteo Barborini for the support provided on the TurboRVB QMC code \cite{TurboRVB}. 
Computational resources were provided by CINECA (Award N. HP10AOW1FU) and by the Caliban-HPC center of Universit\'a dell'Aquila. The authors also acknowledge funding provided by the European Research Council project n. 240624 within the VII Framework Program of the European Union. 


\end{document}